**Metrics matter: Formal comment on Ward *et al Plos-One* paper (2016) "Is decoupling GDP growth from environmental impact possible?"**

Hervé Bercegol[a,b] [*], Paul E. Brockway[c]

*Affiliations*:
a) SPEC, CEA, CNRS, Université Paris-Saclay, CEA Saclay, 91191 Gif-sur-Yvette, France.

b) Laboratoire Interdisciplinaire des Énergies de Demain, Université Paris Cité, CNRS, UMR 8236-LIED, 75013 Paris, France.

c) Sustainability Research Institute, School of Earth and Environment, University of Leeds, Leeds LS2 9JT, UK

* herve.bercegol@cea.fr

*Abstract:*
> The Ward *et al.* (2016) *Plos-One* paper is an important, heavily-cited paper in the decoupling literature. The authors present evidence of 1990-2015 growth in material and energy consumption and GDP at a world level, and for selected countries. They find only relative decoupling has occurred, leading to their central claim that future absolute decoupling is implausible. However, the authors have made two key errors in their collected data: GDP data is in current prices which includes inflation, and their global material use data is the total mass of fossil energy materials. Strictly, GDP data should be in constant prices to allow for its comparison over time, and material inputs to an economy should be the sum of mineral raw materials. Amending for these errors, we find much smaller levels of energy-GDP relative decoupling, and no materials-GDP decoupling at all at a global level. We check these new results by adding data for 1900-1990 to provide a longer time series, and find consistently low (and even no) levels of global relative decoupling of material use. The central claim for materials over the implausibility of future absolute decoupling therefore not only remains valid but is reinforced by the corrected datasets.

## Ward et al. (2016) paper: foundational importance in the decoupling debate

A circular economy in which material resources are kept in the economy as long as possible (to reduce the need for new resources) is often presented as a difficult and necessary goal of future socio-economic transformations. In addition, reducing energy use is one of the core climate change policy goals [1–3], in combination with a rapid switch to renewable energy. The decoupling between GDP and the use of material and energy resources is therefore a basic indicator of progress towards those goals. Thus, the study of decoupling is a hot subject for anyone interested in the sustainability of human civilization. Hence, Ward *et al.* [4] gathered a strong interest, illustrated by 545 citations to date (October 3[rd] 2023) according to the Google Scholar web site.

Unfortunately, two key mistakes in the presentation and use of data affect this work: namely in the historical datasets relating to GDP and minerals use. Due to the foundational importance of the Ward *et al.* [4] paper, it is important to correct the datasets, present and discuss the implications of the updated results. We set out the errors and apply corrections following Ward *et al.'s* [4] sections: their Introduction, and their Australia case study. We additionally provide extended historical data to obtain long-run energy and material decoupling levels from GDP, and finally conclude.

## 1. Corrections to Ward et al.'s Introduction

We introduce two key corrections to the data and results contained in the Introduction. The first correction is to the GDP data. Economic data is usually collected in the local currency that was used in the economic exchanges. However, to compare real values of products and services, this common currency must be adjusted for inflation, all data being converted into a constant currency. Therefore, for use in decoupling studies – i.e. to study the relation between the economic activity and the usage of physical natural resources – one has to use real GDP, expressed in constant currency, as noted by Semieniuk [5], who provides an in-depth treatment of the GDP metric choice. However, it appears in Figure 1 of Ward *et al.* [4] that nominal GDP in current US$ was used for the Introduction's examples of the World, OECD, China and Germany. Indeed, any confusion about their GDP metric choice is clarified by their online journal comment[1] which states that their Figure 1 data uses current GDP in market prices.

The second correction is to material use data, which Ward *et al.* [4] report they take from materials data accessed in 2016 from Lutter *et al.* [6]. We have access to Lutter *et al.* [7] data published online in 2023[2], in which we find that the material use (in tonnes/yr normalized to 1990) shown in their Figure 1 (rising from index of 100 in 1990 to ~150 in 2015 for the World case) is much slower than the sum of the non-metallic and metallic minerals from Lutter *et al.* [7], which is closer to 250 in 2015 (relative to 100 in 1990). Correcting the data means a much higher minerals growth rate is observed.

We present the original and corrected World plots in Figure 1:

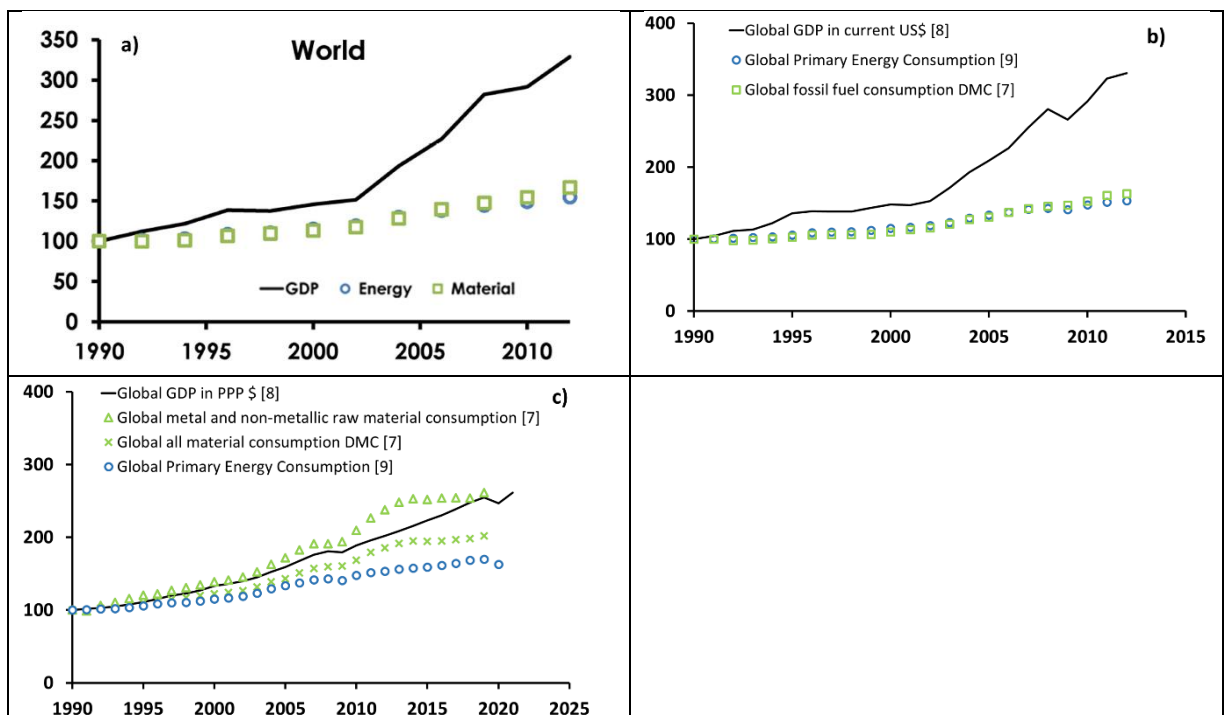

**Figure 1**: comparison of world level data in figure 1 of Ward *et al.* [4] with a reproduction of the figure and a figure obtained with the corrected and adequate data. Scale is normalised to 100 in 1990. **a)** scan of original figure 1 of Ward *et al.* [4] for the world level. **b)** reproduction of a) with data available today for materials [7], GDP [8], and energy [9]. **c)** corrected plot that shows the same energy consumption data as b), but with a series of GDP in constant international currency at Parity of Purchasing Power and corrected materials data of metals and non-metallic material consumption (green triangles), and all

---

[1] https://journals.plos.org/plosone/article/comment?id=10.1371/annotation/3a36e651-ba77-4c92-b3a4-5c2fdc14c34d
[2] Downloading data in 2023 versus 2016 means there may be slight data differences for the same periods. However, we do not observe noticeable changes, as shown in Figure 1a and 1b comparison plots

material consumption (green crosses). All data for materials [7], GDP [8], and energy [9] are obtained from the same sources as Ward *et al.* [4] , and are normalized to 1990 values.

Fig. 1a is a scan of the World plot from Ward *et al.* [4], figure 1; Fig. 1b reproduces 1a with nominal GDP, fossil materials usage data and primary energy consumption series. Figure 1c shows the correct plot with GDP in constant international currency at Parity of Purchasing Power and corrected materials data. Figure 1 uses the same sources [7–9] as Ward *et al.* [4], for GDP, materials and energy. As for materials data, we gathered from Lutter *et al.* [7] metallic and non metallic raw material consumption (green triangles in Fig. 1c) and the total raw material consumption (green crosses in Fig. 1c). The total raw materials include fossil fuels and biomass added to metal and non-metallic minerals. Fig. 1c shows clearly that energy consumption grows slower than real GDP, which is described as relative decoupling [10,11]. It also shows that the metals and non-metallic minerals consumption (green triangles in Fig. 1c) grows slightly quicker than GDP at a global level. When fossil fuels and biomass are added to mineral materials, one obtains a curve intermediary between energy and GDP (green crosses in Fig.1c)

Next, in our Figure 2, we also compare for China the original data of Ward *et al.* [4], in its figure 1 (Fig. 2a) and those reproduced (Fig. 2b) and corrected (Fig. 2c). Whereas the global case of Fig. 1 shows clearly that raw material use has been growing nearly linearly with global GDP on the period, the situation is different for individual countries. In the case of China, raw material consumption (green triangles and crosses in Fig. 2c) decouple from real GDP after 1997. Then the total material use (crosses in Fig.2c) follows energy growth, whereas metallic and non-metallic minerals use (triangles in Fig.2c) grows at an intermediary rate between energy and GDP. One also observes this latter series cease growing after 2013. This peculiar behavior will be discussed further after broaching the case of Australia (see below and Fig. 3).

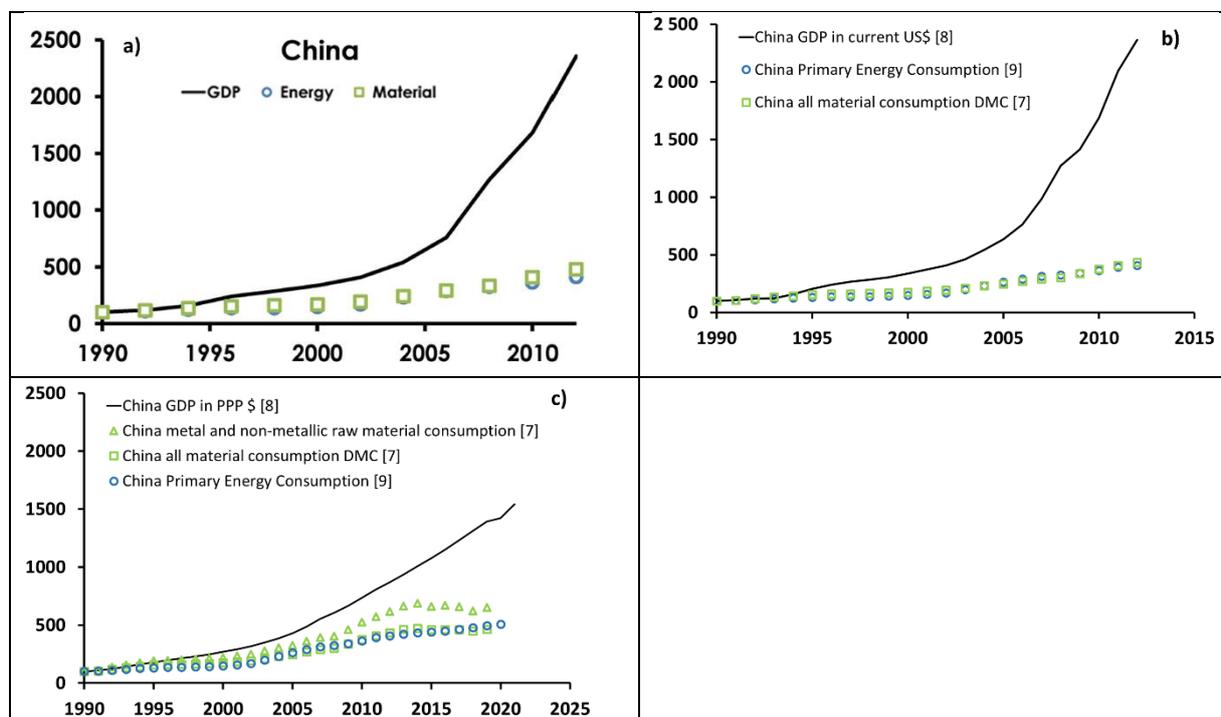

**Figure 2**: comparison of China data in figure 1 of Ward *et al.* [4] with a reproduction of the figure. Scale is normalised to 100 in 1990. **a)** scan of original figure 1 of Ward *et al.* [4] for China. **b)** reproduction of a) with available data today, using the same sources as Ward *et al.* [4] **c)** corrected plot with real GDP, metal and non-metallic minerals use and primary energy consumption (the energy is the same as in b). References indicate sources of the data.

It appears that what is labelled materials in Ward *et al.* [4] plots (Fig 1a and Fig 2a of this work) follows rather closely the primary energy consumption data. In our reproduction in Fig. 1b, the so-called material series seems limited to global fossil fuel consumption. In the case of China, Fig.2b, the material series of Ward *et al.* [4] seems to be the total material use, including energetic materials (fossil fuels and biomass), represented as crosses in Fig. 2b and 2c.

At the world level, there is no indication of any deviation from a linear relation between GDP and mineral material consumption, as seen in Fig. 1c. Previous observations of partial decoupling of mineral use and global GDP was due either to a mistake, as in Ward *et al.* [4], or to a questionable convention. The last case happens when fossil fuels and biomass are mixed with other mineral extracts into a global extraction quantity, as in Krausmann *et al*. [12]. Since energy efficiency increases, the growth rates of energy materials (most of fossil fuels, and some of biomass) and other mineral resources is slower than GDP. In Fig 1c, we have used the sum of metallic and non-metallic raw material consumption, from the same source [7] as Ward *et al.* [4], in the last 30 years, it grew at an average rate greater or equal to that of GDP. If instead of mineral materials, one takes the series of all material usage, including fossil fuels and biomass, as in Fig.1c for the global material use (crosses), one gets a growth rate intermediary between that of energy and that of metals and non-metallic materials.

Overall, from these revised Figures, we observe that constant prices based GDP growth rates are much lower: for the world ~3%/year (versus 5%/year for current prices), and for China ~10%/year (versus 14%/year for current prices) in the period 1990-2012. Second, we see that material use growth rates are higher than Ward *et al.* [4]– we find growth rates for the world of ~3%/year (versus ~2%/year), and for China 7%/year (versus 5%/year). The combined effect (lower GDP growth rate, higher material use growth rate) is to narrow the coupling between resource use and GDP. We still observe (though smaller) relative decoupling for energy-GDP. However, for materials the previous relative decoupling finding has gone at the world level, where we are left with close 1-1 coupling. The tightening of the coupling of global energy and material use with GDP is perhaps nuanced, but is important, as some authors (e.g. [13,14]) refer to Ward *et al.* [4] as a key study of observed relative decoupling in both energy and materials.

## 2. Reappraisal of Ward et al.'s Australia case study

Having set their introductory framing (i.e. only relative decoupling in selected countries and at global level), Ward *et al.* [4] next use data from Hatfield-Dodds *et al.* [15] on Australia's historical energy, materials and GDP, in order to construct a simple model estimate of required material and energy reduction use rates to achieve a state of absolute decoupling for Australia. Reviewing the Hatfield-Dodds *et al.* [15] data, we find the final energy use (in TJ) data appears correct, but questions are raised about the choice of the material series. Moreover, we observe that the behavior of mineral extraction during the last years, which occurred between the publication of Ward *et al.* [4] and the present, evidence the inadequacy of the modelling of Ward *et al.* [4] and Hatfield-Dodds *et al.* [15].

Contrary to material data of the introduction, obtained from Lutter *et al.* [7], Australian data in Ward *et al.* [4] come from Schandl and West [16]. It is clear that the material data in Schandl and West [16] equates to all material consumption: fossil fuels, biomass, metal ores, industrial and construction minerals. We suggest that metal and non metallic minerals (industrial and construction related) should be separated from energy carriers (fossil fuels and biomass). To operate this distinction, we use once again the data of Lutter *et al.* [7], since Hatfield-Dodds *et al.* [15] and Schandl and West [16] do not provide detailed numbers for separate series.

To reproduce Australian data of resource intensities of the economy (figure 2 of Ward *et al.* [4]) we used material and energy consumption series from Lutter *et al.* [7] and bp plc [9]. GDP series obtained from the World Bank [8] were expressed in constant Australian dollars using the constant ratio between Australian dollars 2010 and US$ 2015 calculated by comparing World Bank [8] to Hatfield-

Dodds *et al.* [15] and Schandl and West [16] data. In our Fig. 3, we represent the ratio to real GDP of primary energy consumption in Fig. 3a, of all material extraction series in Fig. 3b and separate material extraction series in Fig 3c.

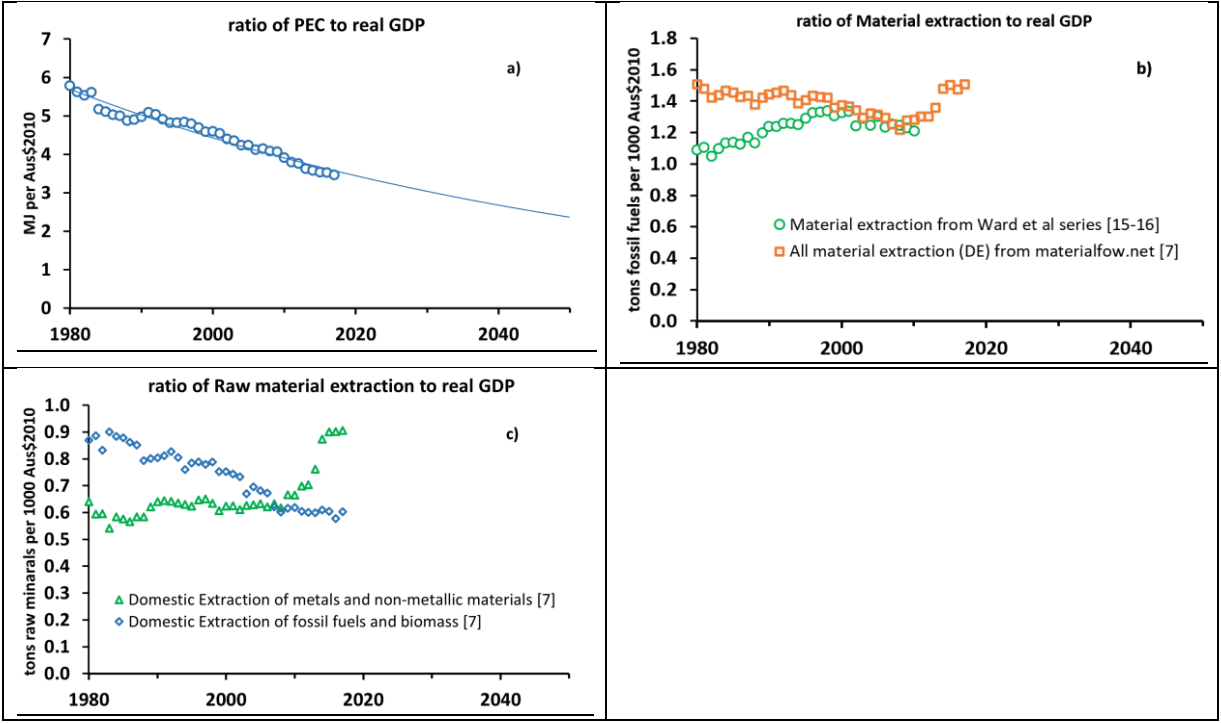

**Figure 3**: evolution of energy and raw materials intensity of the Australian economy. **a)** ratio of primary energy consumption to GDP, with exponential curve fitting. **b)** ratio of all materials extraction to GDP, with data from Ward *et al.* [4] (green circles) and from Lutter *et al.* [7] (orange squares). **c)** ratio of material extraction to GDP, separating Fossil fuel and biomass (blue diamonds) from metals and non mineral materials (green triangles). Data sources for materials [7], GDP [8], and energy [9] are used throughout, except green circles in b) taken from Ward *et al.* [4].

Primary energy usage (Fig. 3a) shows a clear, albeit partial decoupling from GDP, for Australia as for other countries or groups of country shown in figure 1 of Ward *et al.* [4]. In Fig. 3b, we compare the Australian material intensity data of Ward *et al.* [4], with our own values calculated using available data for materials [7], GDP [8], and energy [9]. We observe that the series differ noticeably in the period 1980-2000 but are similar in 2000-2010. At this point we cannot explain the discrepancy during the older period (it might come from different conventions…) : we do not comment on these features. What seems more relevant to decoupling is the sudden jump of material intensity in the later period: from 2012 to 2015, the ratio grows from 1.2 ton per 1000 Aus$ 2010 to 1.5 ton. This kind of behavior cannot be rendered by the exponential curve fitting used in Ward *et al.* [4] and is not envisaged in scenarios of Hatfield-Dodds *et al.* [15]. Moreover, when materials with mainly an energetic use (fossil fuels and biomass) are separated from metal and non-metallic minerals as in Fig. 3c, one sees that the sudden increase of the material intensity is totally due to a similar increase (+0.3 ton/1000 Aus$ 2010) in the extraction of the second type. To our point of view thus, a more relevant series to study the evolution of the economic intensity of material use is that of the extraction of metals and non-metallic material, excluding energetic materials (fossil fuels and biomass).

Coming back to the case of material consumption in China (green triangles in Fig. 2c): the leveling of the curve after 2013 should not be interpreted as a complete decoupling of Chinese GDP from raw materials consumption after 2013. Indeed, knowing the entanglement (imbrication) of national outputs in the global economy, such a statement is totally unwarranted, especially when one sees in Fig 3c the concomitant increase of raw material intensity in Australia at the same time. Nevertheless, these interesting features are certainly worth some economic analyses, which are beyond the scope of this comment.

### 3. Extension of historical global data supports the no decoupling hypothesis

Finally, to test the corrected Introduction results for Ward *et al.* [4] but over a longer timeseries, we add data for 1900-1990. Fig 4 establishes more firmly the linear relation between global GDP and usage of mineral material resources. Using data made available with the publications of Krausmann *et al*. [12,17] these graphs show that annual consumption mineral materials all follow closely the pace of global GDP. It is both true for the last 30 years (Fig. 4a), and from the beginning of the 20[th] century (Fig. 4b). When all national contributions add-up, one get the behavior of Fig. 4: the consumption of mineral raw materials grows proportionally to GDP and has been doing so for the last hundred years.

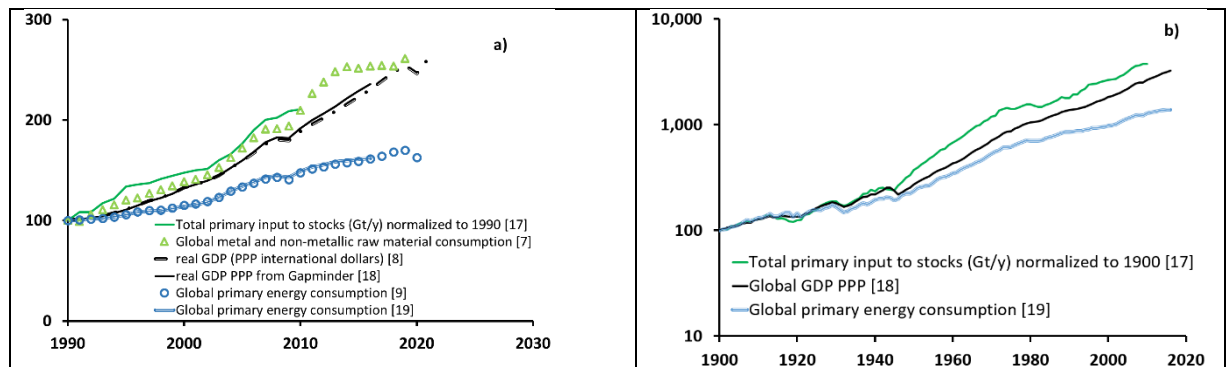

**Figure 4**: comparison of data for global GDP [8,18], Primary Energy Consumption [9,19] and raw material consumption [7,17]. **a)** from 1990, normalized to 1990 values. **b)** from 1900, normalized to 1900 values, semi-log scale. References indicate sources of the data.

At a global level, GDP grew at an average 3% annual pace for the last hundred years, whereas energy consumption showed an average 2% annual growth rate [11], indicating thus a partial-only decoupling. The so-called energy rebound effect(s) is a key candidate to explain the inexistence of absolute decoupling in available economic data [10].

### 4. Conclusion

The importance of using the correct data in decoupling studies is paramount, and researcher need to pay close attention to their datasets. Here we found in an influential paper that GDP and material use data was incorrect. Overall, the corrected data means much lower levels of global relative decoupling of material and energy use – something we reinforce by adding data for 1900-1990 to provide a longer time series. Indeed, our corrections and extensions add further weight to their central claim - of the implausibility of future absolute decoupling under historical GDP growth rates.

### 5. Declaration of competing interest

The authors declare that they have no known competing financial interests or personal relationships that could have appeared to influence the work reported in this paper.

### 6. Data Availability

All relevant data are within the paper and its Supporting Information files.

## 7. Acknowledgements

Paul Brockway's time was funded by the UK Research and Innovation (UKRI) Council, supported under Engineering and Physical Sciences Research Council Fellowship award EP/R024254/1.

## 8. CRediT author contributions

HB: Conceptualization, Methodology, Validation, Formal analysis, Investigation, Data Curation, Writing - Original Draft, Writing - Review & Editing, Visualization. PB: Conceptualization, Methodology, Investigation, Writing - Review & Editing, Funding acquisition.